# Double-containment coil with enhanced winding mounting for transcranial magnetic stimulation with reduced acoustic noise

Lari M. Koponen, Stefan M. Goetz, *Member, IEEE*, and Angel V. Peterchev [*], *Senior Member, IEEE*

*Abstract*—*Objective:* This work aims to reduce the acoustic noise level of transcranial magnetic stimulation (TMS) coils. TMS requires high currents (several thousand amperes) to be pulsed through the coil, which generates a loud acoustic impulse whose peak sound pressure level (SPL) can exceed 130 dB(Z). This sound poses a risk to hearing and elicits unwanted neural activation of auditory brain circuits. *Methods:* We propose a new double-containment coil with enhanced winding mounting (DCC), which utilizes acoustic impedance mismatch to contain and dissipate the impulsive sound within an air-tight outer casing. The coil winding is potted in a rigid block, which is mounted to the outer casing by its acoustic nodes that are subject to minimum vibration during the pulse. The rest of the winding block is isolated from the casing by an air gap, and sound is absorbed by foam within the casing. The casing thickness under the winding center is minimized to maximize the coil electric field output. *Results:* Compared to commercial figure-of-eight TMS coils, the DCC prototype has 10–33 dB(Z) lower SPL at matched stimulation strength, whilst providing 22% higher maximum stimulation strength than equally focal commercial coils. *Conclusion:* The DCC design greatly reduces the acoustic noise of TMS while increasing the achievable stimulation strength. *Significance:* The acoustic noise reduction from our coil design is comparable to that provided by typical hearing protection devices. This coil design approach can enhance hearing safety and reduce auditory co-activations in the brain and other detrimental effects of TMS sound.

## I. INTRODUCTION

TRANSCRANIAL magnetic stimulation (TMS) is a noninvasive method for brain stimulation, with both clinical and research applications. In TMS, an electromagnet coil placed on the subject's scalp is pulsed to create a rapidly changing magnetic field (B-field) which induces an electric field (E-field) in the vicinity of the coil. A typical biphasic TMS pulse lasts only about 300 μs, but must produce peak magnetic field on the order of 1 T, which requires a coil current over 1,000 A. The high current and magnetic field produce a mechanical vibration of the coil, which manifests itself in a loud, impulsive sound with peak sound pressure levels (SPL) close to 140 dB(Z) [1]. For pulse trains during repetitive TMS (rTMS), the continuous sound level (SL) can exceed 110 dB(A) [1].

The coil sound is a significant limitation of TMS. It poses a risk to hearing [1]–[4] and, with missing or inadequate hearing protection, can cause permanent hearing damage [5]. For some rTMS protocols and devices, the sound level may be high enough to indicate hearing protection for device operators near the TMS coil as well [1]. Further, loud sounds can contribute to headaches [6], [7], which are a common adverse side effect of TMS [2]. Loud sounds may also be disturbing to subjects with autism spectrum disorder [8], [9] or post-traumatic stress disorder [10]. Generally, the impulsive noise from TMS devices propagates beyond the room where the device is operated and can therefore be disruptive in clinical or research settings. Moreover, the pulse sound reduces the effective focality of TMS since auditory pathways are activated synchronously with the electromagnetic stimulation of the targeted cortical region. These parallel activation paths are hard or even impossible to separate, for example, in neuroimaging data [11], [12]. Finally, during rTMS, the acoustic stimulation may cause unwanted neuromodulation [13], [14].

There are several adopted or proposed approaches to mitigate the effects of the TMS sound. Adequate hearing protection during TMS can be obtained with either earmuffs (typical attenuation 20–30 dB for relevant frequencies, i.e., above 1 kHz [15]) or correctly worn earplugs (typical attenuation 20–25 dB for the same frequencies [15]). Indeed, correctly applied earplugs appear to ensure hearing safety in TMS [2], [4], [16], [17]. Teaching proper insertion technique to subjects helps [18], but a consistently good fit of earplugs can be challenging to

This is an unrefereed preprint. Research reported in this publication was supported by the National Institute of Mental Health of the National Institutes of Health under Award Number R01MH111865 as well as by hardware donations from Magstim. The content is solely the responsibility of the authors and does not necessarily represent the official views of the National Institutes of Health.

L. M. Koponen is with Department of Psychiatry & Behavioral Sciences, Duke University, Durham, NC, USA.

S. M. Goetz is with Department of Psychiatry & Behavioral Sciences, Department of Electrical & Computer Engineering, and Department of Neurosurgery, Duke University, Durham, NC, USA.

*A. V. Peterchev is with Department of Psychiatry & Behavioral Sciences, Department of Electrical & Computer Engineering, Department of Neurosurgery, and Department of Biomedical Engineering, Duke University, Durham, NC, USA. (correspondence e-mail: angel.peterchev@duke.edu).

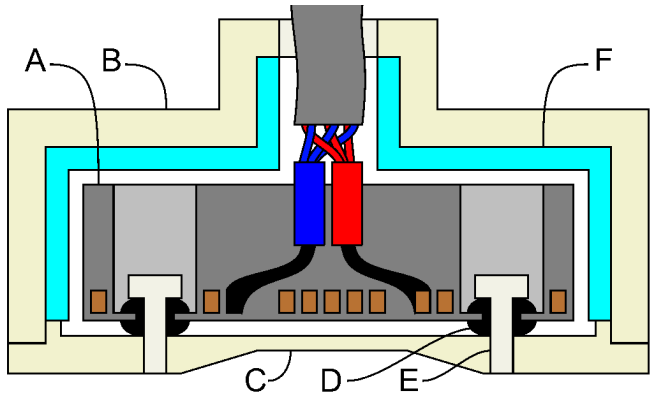

Fig. 1. The double-containment coil comprises a winding block (A), which is essentially a fully-fledged TMS coil, and an outer casing (B) with a lid with a central recession to reduce the winding-to-head distance (C). The winding block is mounted to the outer casing with rubber grommets (D) and nylon bolts (E) at the points of minimal in-plane vibration. The outer casing walls not facing the head are further covered with thin acoustic foam panels (F) to reduce reverberation.

obtain for all subjects [18]–[21]. Indeed, potential permanent hearing damage from TMS was observed, likely due to incorrectly applied ear plugs [5]. Beyond hearing protection devices, the perceived sound can be reduced by inserting a layer of foam between the coil and the scalp to decrease bone-conduction of the sound [11], [22]. However, this added distance between the winding and the brain reduces both the energy efficiency and attainable stimulation focality—if the coil windings are not optimized for the extra spacing, the efficiency loss is about 10% per mm [23]. The windings of typical commercial coils are 2–5 mm from the surface, but MRI-specific TMS coils may have up to 10 mm of extra sound-absorbing acoustic foam, which reduces the maximum output even when the coil windings are made larger for less focal stimulation [1], [24].

In principle, the sound intensity reaching the cochlea could also be reduced with active noise cancellation (ANC) technology. Conventional real-time ANC solutions, however, are typically limited to steady-state sounds and lower frequencies, providing attenuation only for frequencies below 1 kHz, even with in-ear headphones and for sound intensities much lower than TMS [25], [26]. The TMS coil click has peak SPL that would require extremely powerful headphones, and contains mostly frequencies that are too high for ANC. A TMS-specific offline ANC solution could theoretically solve the problem with high frequencies, but even in simulations, the attenuation for frequencies above 1 kHz was rendered close to zero with a small change in the coil orientation [27]. An ANC solution would also not reduce the bone-conducted sound. Importantly, none of the approaches described so far is sufficient to prevent auditory brain activation. Consequently, noise played through earphones, e.g. fixed 90 dB(A) or individually leveled white noise, is sometimes used to mask the TMS sound [28], [29]. By practically raising the hearing threshold, such noise masking can reduce unwanted TMS-synchronized auditory activation. However, the loud masking sound itself may disturb noise-sensitive subjects and patients;

hinder verbal communication, auditory tasks, or psychotherapy during the TMS session; reduce cognitive performance [30]; and require noise dosimetry to ensure adhering to hearing safety limits [31], [32].

Considering the adverse impact of the loud TMS sound and the limitations of mitigation approaches, it is compelling to develop TMS devices with lower acoustic emission. This approach is further supported by the conventional hierarchy of hazard controls, in which personal protective equipment is considered the least effective, last-resort solution [33]. We have proposed a two-pronged approach to quiet TMS, involving improved electromechanical coil design and briefer pulses [34], [35]. In the present work we focus on the first part of this approach and demonstrate a TMS coil design with high electromagnetic output but substantially reduced acoustic emission for conventional TMS pulse waveforms.

## II. Materials and methods

### A. Coil structure

The proposed coil design is diagrammed in Fig. 1. The design has a double-containment structure, in which a potted optimized winding is contained within an independent outer casing, separated from the head-facing side of the winding block by a 1.6 mm air gap. Further, to minimize the distance to the windings while retaining structural rigidity, the head-facing side of the outer casing (lid) incorporates at its center a 120 mm diameter circular recession tapering down to a 75 mm diameter flat section with a thickness of only 1.5 mm. To minimize the sound transmission via the mounting points for the winding block, their locations were optimized to coincide with nodal points of minimum in-plane vibration of the winding block, determined from an electromechanical simulation. The mounting points are equipped with commercial styrene-butadiene rubber grommets to reduce further the transmission of mechanical vibrations to the coil lid.

For our prototype, the lid was 3d printed with selective laser sintering from a 30% glass-filled nylon 12 (Xometry, USA), whereas the rest of the outer casing was built around a prefabricated acrylic box (Fig. 2, right). The two parts were connected by bolts, and the interface was sealed with a custom laser-cut butyl rubber gasket. The coil windings of the prototype were wound from a 4.67 mm × 2.69 mm rectangular litz wire (120 strands of 0.2 mm enameled copper wire, total copper cross section of 3.84 mm$^2$) with a 0.38 mm fluorinated ethylene propylene jacket (New England Wire, USA). The winding block was constructed by potting the wire windings with corundum-filled high-strength lamination epoxy (Fibre Glast, USA) (Fig. 2, middle). The potting mold was 3d-printed from nylon 12 (Xometry, USA), and had a minimum wall thickness of 0.7 mm and minimum potting thickness of 0.8 mm (Fig. 2, left). Consequently, the coil windings were 1.5 mm above the bottom of the winding block, and the total distance between the center of the coil windings and the coil surface was 6.9 mm, which is comparable with commercial TMS coils [34].

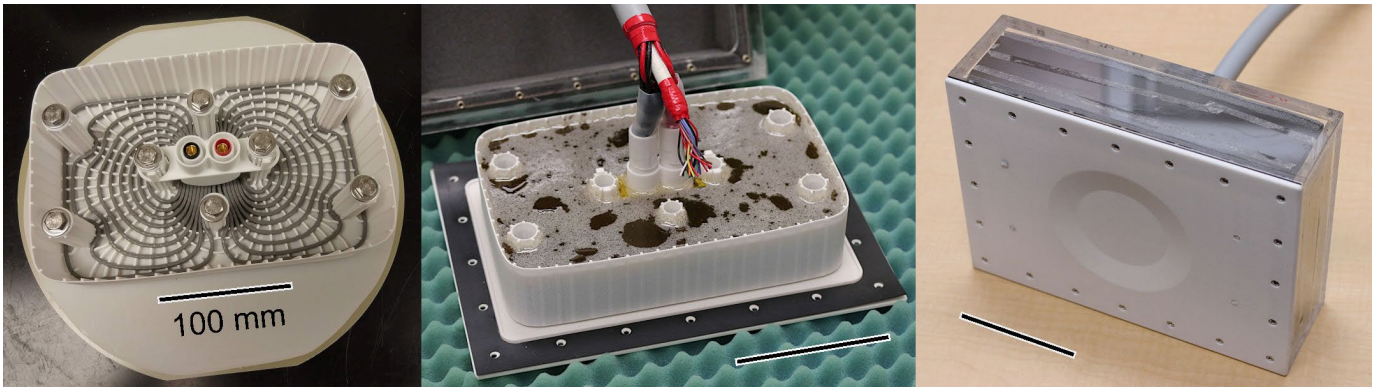

Fig. 2. The double-containment coil prototype is implemented with energy-optimized litz-wire windings in a 3d-printed sintered nylon mold (left), potted in place with corundum-filled epoxy (middle), and contained in an outer casing constructed from acrylic plastic and glass-filled nylon (right). The scale bar in each pane is 100 mm long.

The winding was connected to a commercial TMS device (MagPro X100 incl. MagOption, MagVenture, Denmark) with a 3 m low-inductance TMS-coil cable (Magstim, UK) and a customized orange-type SBE 160 power connector (Anderson Power Products / Ideal Industries, USA). The cable exit from the outer casing was sealed with an air-tight cord grip, which was separated from the rest of the outer casing with a butyl rubber gasket.

### B. Coil winding optimization

The optimization problem for the energy efficiency of TMS coil windings is a convex optimization problem [36]. Such problems have a somewhat shallow energy landscape around the optimum. Thus, minor sacrifices in efficiency can lend improved buildability and desired electrical properties such as higher inductance for a given number of turns with lower coil current requirements. We solved this problem with TMS-coil optimization routines further developed from our prior work [23]. Specifically, we added two new types of constraints: a constraint for the magnitude of coil current density and for the maximum dI/dt for the desired electric field (E-field) in the cortex. The former is a constraint for a norm, solved similarly to the previous E-field norm constraints [23] and satisfied to a tolerance of 0.001. The updated optimization routines were implemented with MATLAB (Global Optimization Toolkit, Version R2018a, Mathworks, USA).

### C. Acoustic simulations

For acoustic simulation of the coil winding block, we built two models. First, a simple 2d model for the in-plane vibrations was used to tune the optimization constraints for the coil windings. Second, a detailed 3d model was created to estimate the required thickness for the windings block. The latter model was further validated post-hoc against the acoustic measurements. For the models, the material parameters for the corundum-filled epoxy were estimated with the S-combining rule [37]. Both models were solved with COMSOL Multiphysics (Version 5.3a, COMSOL, USA).

### D. Electrical simulations

A specific TMS coil design has three key electrical parameters: the inductance and resistance of the windings as well as the coupling coefficient to the brain, defined as the ratio between the strength of the E-field induced in the cortex and the rate of change of the coil current. We computed the coupling coefficient for a 85 mm spherical head model [38] with the triangle construction [24], [39] implemented in Mathematica (Version 12.0.0.0, Wolfram Research, USA). The coil inductance and resistance were computed with multipole-accelerated inductance extraction [40] (FastHenry2, Software Bundle 5.2.0, FastFieldSolvers, Italy), and the power cable contribution was modelled with COMSOL.

### E. Acoustic and electrical measurements

The acoustic and electrical measurements of the coil were carried out similarly to our previous work characterizing commercial TMS coils [1] with a few minor differences. Notably, we omitted the use of a soundproof chamber and measured the sound in a regular TMS treatment room, since we focused on near- and supra-threshold pulses which produce sound significantly above the ambient noise level.

Briefly, for acoustic measurements, an omnidirectional flat-frequency-response pressure microphone (M50, Earthworks Audio, USA) was placed 25 cm below the center of the head-facing side of the coil. The microphone output was fed to a wide-input-signal-range preamplifier (RNP8380, FMR Audio, USA) and then an audio interface with a sample rate of 192 kHz (U-Phoria UMC404HD, Behringer, Germany). The measurement system was calibrated with a 1 kHz, 1 Pa reference sound pressure source (407722, Extech Instruments, USA). We recorded the sound from single TMS pulses at 10% to 100% of maximum stimulator output (MSO) in 10% MSO increments. The continuous sound of rTMS was synthesized from these pulses. To extract the SPL and SL, the audio was processed with MATLAB Audio Toolbox. We used the electromagnetic artefact removal algorithm as well as low- and high-pass filters described in our previous study [1].

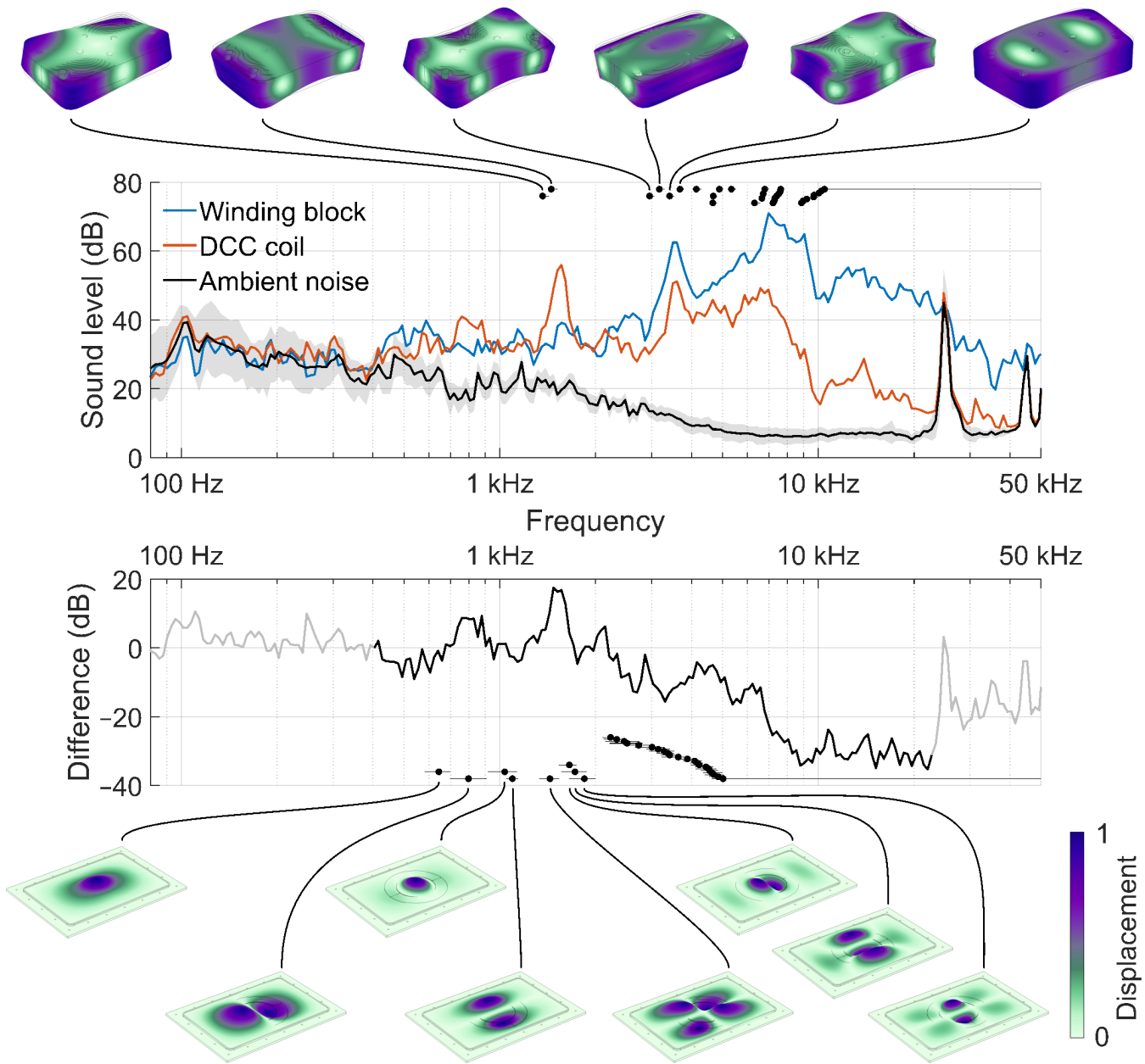

Fig. 3. Measured sound spectra and simulated mechanical vibration modes of the prototype double-containment coil (DCC). Top: The six lowest vibration modes of the winding block linked to the respective spectral frequencies. The whiskers denote the model uncertainty arising from uncertainty in the material parameters and winding block thickness. Second row: The 1/24-octave sound level of the complete coil and its winding block at 167% average resting motor threshold (RMT). Third row: The attenuation provided by the outer casing obtained by subtracting the winding block spectrum from the complete coil spectrum. The attenuation spectrum at frequencies below 500 Hz and above 25 kHz could not be measured reliably and is therefore grayed out. Bottom: The seven lowest vibration modes of the outer casing lid. The whiskers denote the model uncertainty arising from uncertainty in the stiffness of the coupling to winding block and the thickness of the lid recession.

The measurement distance, 25 cm, was chosen to avoid inadequate sampling of the sound in the near field and allow filtering out the electromagnetic artefact from the stimulation [1]. As the sound of TMS attenuates inversely with distance for distances down to about 5 cm [41], the SPL and SL at the approximate location of the subject's ears, 5 cm from the coil, can be estimated by adding 14 dB to these results [1].

The induced E-field was measured with a printed-circuit-board-based triangular probe [1] connected to an oscilloscope (DS1052E, Rigol, China) with a sampling rate of 250 MHz. To estimate the effective neural stimulation strength, the recorded waveform was fed into a strength–duration model [42], [43] with a time constant of 200 μs. Additionally, we recorded the maximum rate of change for the coil current from the sensor built into the TMS device. The stimulation strength was calibrated to the average measured resting motor threshold (RMT) of normal subjects extracted from the literature [1].

## III. Results

### A. Coil windings and construction

The acoustic simulations of the in-plane vibrations of the windings gave up to four nodal points of greatly reduced mechanical vibrations. The locations of these points depend mostly on the coil size, and to lesser extent on Poisson's ratio of the potting material. For epoxy-like materials (Poisson's ratio about 0.3), four nodal points were identified in the corners of a 180 mm × 130 mm winding block. To move these points away from the corners and place them along the nodal line for the lowest resonant mode of the winding block, we chose to implement a slightly larger, 225 mm × 145 mm winding block. To obtain adequate stiffness and sufficiently high resonant frequencies for the out-of-plane vibration modes, the out-of-plane vibration model suggested winding block thickness of at least 40 mm; therefore, we chose a thickness of 45 mm. We designed the coil to match the E-field focality of a Magstim 70mm Double Coil in the 85 mm spherical head model. The resulting windings are shown in Fig. 2 (left).

For the potting material, we chose an epoxy-to-corundum mass mixing ratio of 1:2 (35.8% corundum by volume), which was the highest fill ratio with adequate fluidity for an easy pour. This resulted in a suboptimal structure of the winding block, as the filler sedimented before the mixture cured. The realized thickness of the potting was 43.3 mm (Fig. 2, middle). The resulting winding block has an apparent line that separates the bottom with corundum-filled epoxy (31.5 mm) from the top with essentially neat epoxy (11.8 mm). Assuming negligible amount of filler in the top layer, the resulting post-sedimentation fill factor for the bottom layer is about 52.0% by volume, which is close to the maximum mixing ratio obtained during earlier prototyping (1:4 mass ratio, 52.7% corundum by volume).

### B. Electrical properties

The simulated coil inductance and resistance were, respectively, 11.0 μH (10.8 μH for the coil windings and 0.15 μH for the power cable) and 25.7 mΩ (20.2 mΩ for the windings and 5.3 mΩ for the cable). These values matched very well the respective measurements of 11.5 μH and 27.0 mΩ at 10 kHz, acquired with B&K Precision Model 889A Bench LCR/ESR Meter (B&K Precision Corporation, USA). The unaccounted inductance and resistance likely stem from parasitic inductance and resistance associated with the connections between the winding, coil cable, TMS device, and measurement probe.

The simulated coupling coefficient to cortex was 1.46 (V/m)/(A/μs) for the entire coil, and 1.65 (V/m)/(A/μs) for the exposed coil winding block. When connected to the MagPro TMS device, the pulse duration for biphasic TMS pulses was 299 μs, which was close to conventional MagVenture coils. The measured coupling coefficients were 1.44 (V/m)/(A/μs) for the coil, and 1.60 (V/m)/(A/μs) for the exposed winding block, agreeing well with the simulations. Thus, the outer casing reduced the E-field magnitude and the associated stimulation strength by 10%. Consequently, the stimulation strength at 100% MSO was 263% and 293% of average RMT for the entire coil and the exposed winding block, respectively.

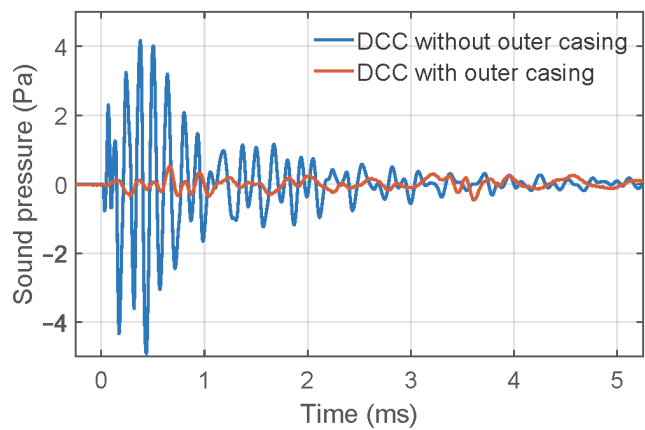

Fig. 4. Sound pressure waveforms from the double-containment coil prototype (DCC). The start of the TMS pulses is at –0.73 ms to compensate for the sound propagation delay in the air. The exposed winding block (108 dB(Z) peak) is compared to the complete coil with outer casing (88 dB(Z) peak). Both configurations are measured for stimulation strength of 167% RMT; thus, the complete coil had 11% higher current to compensate for the thickness of the casing.

### C. Acoustic properties

The SL of the ambient noise in our TMS treatment room was 45 dB(A), and the peak SPL in the 0.2 s measurement window was 71 dB(Z), both about 25 dB above the ambient noise in our earlier measurements inside a soundproof chamber [1]. Given the reduction of SL and SPL for the DCC compared to commercial TMS coils, the ambient noise prevented measuring the sound from subthreshold pulses, but was low enough to have negligible effect on the sound recordings from near- and supra-threshold pulses. In addition to the elevated noise background, we further identified a few narrowband ultrasonic sound sources, at 25.0, 45.1, and 51.5 kHz, likely from presence sensors for the room lighting and air conditioning. The strongest of these three sources was at 25.0 kHz and had 1/3-octave sound level of 35 dB. Given their intensities, these too had negligible effect on the SL and SPL for suprathreshold TMS (see Fig. 3).

As the coil sound scales similarly to other air-core TMS coils, we report numbers only for a stimulation strength of 120% RMT for a subject with a top 5 percentile RMT, i.e., about 167% of average RMT [1]. For rTMS, we used the highest repetition rate sustained for several seconds in clinical treatments, 20 Hz [44], [45]. These numbers can be scaled to other stimulation strengths and repetition rates as described in [1].

For the coil winding block, the peak SPL at 167% RMT was 108 dB(Z). With the outer casing, the peak SPL was reduced by 20 dB(Z) to 88 dB(Z) (see Fig. 4). With C-weighting, which filters out higher frequencies, the peak SPLs were 105 dB(C) and 87 dB(C), respectively. The peak SPL was 10 dB(Z) lower than the quietest coil in our database [1], which is a commercial MRI-compatible coil (MagVenture MRI-B91); 16 dB(Z) lower than the quietest conventional TMS coil; 23 dB(Z) lower than the only coil with a comparable maximum stimulation strength, which has an angled winding topology; and 33 dB(Z) lower than the loudest coil (Fig. 5, top). The reductions in C-weighted sound were similar.

The continuous SL of a 20 Hz rTMS train, for the coil winding block, was 83 dB(A). With the outer casing this level was reduced by 14 dB(A) to 69 dB(A). This was 7 dB(A) lower than the commercial MRI-compatible coil, 11 dB(A) lower than the best conventional coil, 16 dB(A) lower than the only coil with comparable maximum stimulation strength, and 26 dB(A) lower than the loudest coil (Fig. 5, bottom).

The 1/24-octave sound spectrums (Fig. 3, second row) indicate that the winding block emits most of its sound around 7 kHz, i.e., at twice the TMS pulse frequency of 3.35 kHz. This is expected for normal TMS coils, as the mechanical vibrations are driven by the Lorentz forces which are proportional to the squared coil current, and hence have their spectral power peak at double the current frequency. In addition, there is a visible harmonic around 14 kHz and two visible subharmonics near 3.5 kHz and 1.6 kHz (corresponding closely to the lowest two sets of vibration modes for the windings block, illustrated in Fig. 3, top). With the outer casing, a new resonant peak is formed around 0.8 kHz, the peak at 1.6 kHz is amplified, the remaining peaks at 3.5, 7, and 14 kHz are attenuated, and there is minimal amount of near-ultrasound content. The peaks at 31 and 38 kHz likely originate from the power electronics within the TMS device. Thus, the outer casing of the coil acts as an acoustic low-pass filter, which provides about 30 dB attenuation for frequencies above 8 kHz, whilst amplifying sound around some of the lowest vibration modes (Fig. 3, bottom). The amplification at 1.6 kHz is likely due to the coupled oscillations between the modes for the lid and its thin window.

## IV. Discussion

We presented a new coil design to reduce the sound of TMS. This double-containment coil design (DCC) maximized the mismatch in acoustic impedance in the path between the winding and the casing [34] without increasing significantly the thickness of the acoustic containment structure. This is in contrast to previously suggested TMS sound containments utilizing medium to high vacuum of below 1 Pa [46]. The proposed sound containment provides superior acoustic insulation compared to a layer of acoustic foam that is approximately twice as thick in commercial MRI-compatible TMS coils, which are relatively quiet but have reduced maximum stimulation strength. Our coil prototype further utilized windings that were optimized for maximal energy efficiency despite the additional thickness of the casing. This resulted in a coil that, with the same TMS device, has both higher maximum stimulation strength and lower acoustic emissions than any conventional flat figure-8 coil we tested.

The DCC was designed to be compatible with ultra-brief TMS pulses, where the total duration of the biphasic pulse will be reduced by an order of magnitude, from 300 µs down to about 30 µs [35]. These pulses are expected to require comparable peak currents, which necessitates next-generation

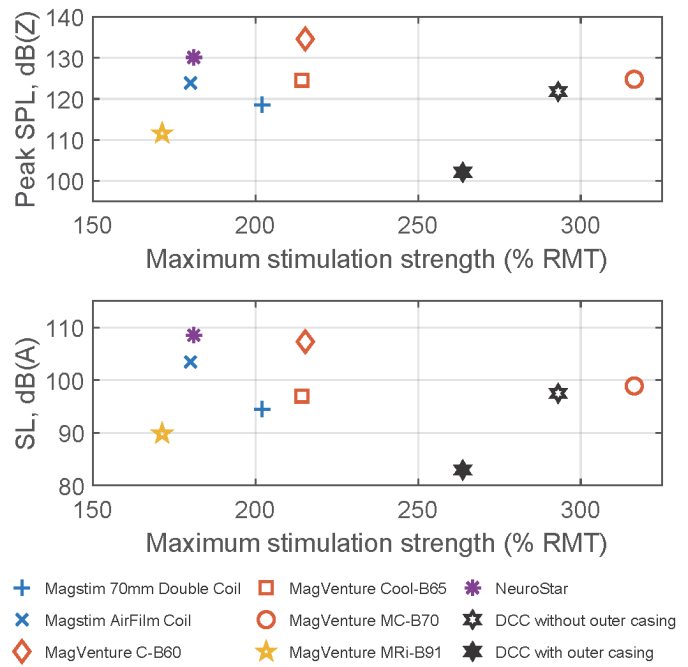

Figure 5. Measured sound levels of various coils at a matched stimulation strength as a function of the maximum stimulation strength obtained at maximum stimulator output. Top: peak SPL at 167% average RMT at 5 cm from coil. Bottom: SL of 20 Hz rTMS at 167% average RMT at 5 cm from coil. Apart from the DCC measurements, the commercial coil data are reproduced from our prior work [1].

TMS devices with operating voltages in excess of 10 kV [35]. Such ultra-brief pulses are expected to further reduce the sound level by both pushing the spectral peak at two times the pulse current frequency outside the human hearing range and reducing the energy in the excited subharmonics by a factor of 10. Given that the DCC containment provides better attenuation at higher frequencies, these two combined may further reduce both SPL and SL by more than 20 dB for a truly quiet TMS device. Even without these future developments, the present coil prototype provides superior performance to existing TMS coils with the conventional pulse waveforms used in this work. Notably, the DCC prototype has 20 Hz rTMS sound level reaching the subject that is below the lowest standard hearing safety limit (85 dB(A) for more than 1 s [32]).

The prototype DCC has some design and construction limitations that could be addressed in the future to improve performance further. For example, we chose to use jacketed litz wire for our windings to ensure compatibility with the higher operating voltages required for suprathreshold ultra-brief pulses [35]. For conventional TMS pulses, which do not use such higher voltages and current frequencies, the space taken by the flexible jacket material can be replaced with either more potting compound or copper, and the litz wire may be replaced with solid rectangular magnet wire, both of which increase stiffness and thus reduce the emitted sound. Similarly, the winding block has four additional holes near its center (Fig. 2, left) which our earlier outer-casing prototypes used for support rods for the lid. These holes impose an additional design constraint for the windings, reducing the efficiency by about 3%. As these holes are not needed in the present design, omitting them would increase the maximum stimulation strength, and consequently slightly decrease the sound at matched output. Further, in the prototype, the filler material in our potting compound sedimented on the bottom, leaving the top essentially devoid of the filler material (Fig. 2, middle). Compared to an ideal potting, this reduced the stiffness of the top part of the winding block, making the winding block louder. Finally, the outer containment of the prototype is both larger and sturdier than necessary as it was built from a readily available prefabricated acrylic box. The dimensions of the outer casing can be reduced, and its walls could be made lighter with no adverse effect on the sound. In addition, a better coil could be built by replacing some of the simple materials used in the prototype with more advanced materials, like constructing the lid from a fiberglass-reinforced plastic instead of glass-filled plastic, or milling parts of the winding block from either fiberglass or machinable ceramics, which are 2 to 7 times stiffer than the corundum-filled epoxy.

Finally, some aspects of the DCC prototype were designed based on qualitative considerations and approximations, rather than measured properties of the complete prototype. For example, we chose to implement the thinnest practical combination of air gap and lid (3.1 mm), as it best highlights the design concept and gives us largest possible headroom in stimulation strength with ultra-brief TMS pulses whose stimulation efficiency is presently unknown [35]. Should this headroom not be needed, the sound attenuation by the outer casing can be improved by increasing either the width of the airgap (which reduces the duration of the sound reverberation inside the outer containment) or the thickness of the window in the lid (which further reduces the sound transmission). The optimum ratio between the two depends on both the lid material and the desired total thickness.

## V. Conclusion

The proposed DCC coil design substantially reduces the instantaneous peak sound pressure level and the continuous sound level during TMS, while providing stronger maximum stimulation output. This can mitigate problems associated with the TMS coil sound.

## Declaration of competing interest

L. M. Koponen, S. M. Goetz, and A. V. Peterchev are inventors on patents and patent applications on TMS technology including the quiet TMS coil technology described in this paper. S. M. Goetz has received research funding from Magstim Inc. related to TMS technology, A. V. Peterchev has received research and travel support as well as patent royalties from Rogue Research; research and travel support, consulting fees, as well as equipment donation from Tal Medical/Neurex; patent application and research support from Magstim; equipment loans and hardware donations from MagVenture; and expert witness compensation from Neuronetics.

## References

[1] L. M. Koponen, S. M. Goetz, D. L. Tucci, and A. V. Peterchev, "Sound comparison of seven TMS coils at matched stimulation strength," *Brain Stimulat.*, vol. 13, no. 3, pp. 873–880, May 2020, doi: 10.1016/j.brs.2020.03.004.

[2] S. Rossi, M. Hallett, P. M. Rossini, and A. Pascual-Leone, "Safety, ethical considerations, and application guidelines for the use of transcranial magnetic stimulation in clinical practice and research," *Clin. Neurophysiol.*, vol. 120, no. 12, pp. 2008–2039, Dec. 2009, doi: 10.1016/j.clinph.2009.08.016.

[3] S. M. Goetz, S. H. Lisanby, D. L. K. Murphy, R. J. Price, G. O'Grady, and A. V. Peterchev, "Impulse noise of transcranial magnetic stimulation: measurement, safety, and auditory neuromodulation," *Brain Stimulat.*, vol. 8, no. 1, pp. 161–163, Jan. 2015, doi: 10.1016/j.brs.2014.10.010.

[4] R. L. Folmer and S. M. Theodoroff, "Hearing protective devices should be used by recipients of repetitive transcranial magnetic stimulation," *J. Clin. Neurophysiol.*, vol. 34, no. 6, p. 552, Nov. 2017, doi: 10.1097/WNP.0000000000000413.

[5] A. Zangen, Y. Roth, B. Voller, and M. Hallett, "Transcranial magnetic stimulation of deep brain regions: evidence for efficacy of the H-coil," *Clin. Neurophysiol.*, vol. 116, no. 4, pp. 775–779, Apr. 2005, doi: 10.1016/j.clinph.2004.11.008.

[6] P. R. Martin, J. Reece, and M. Forsyth, "Noise as a trigger for headaches: relationship between exposure and sensitivity," *Headache J. Head Face Pain*, vol. 46, no. 6, pp. 962–972, May 2006, doi: 10.1111/j.1526-4610.2006.00468.x.

[7] C. Wöber, J. Holzhammer, J. Zeitlhofer, P. Wessely, and C. Wöber-Bingöl, "Trigger factors of migraine and tension-type headache: experience and knowledge of the patients," *J. Headache Pain*, vol. 7, no. 4, pp. 188–195, Sep. 2006, doi: 10.1007/s10194-006-0305-3.

[8] S. Khalfa *et al.*, "Increased perception of loudness in autism," *Hear. Res.*, vol. 198, no. 1, pp. 87–92, Dec. 2004, doi: 10.1016/j.heares.2004.07.006.

[9] L. N. Stiegler and R. Davis, "Understanding sound sensitivity in individuals with autism spectrum disorders," *Focus Autism Dev. Disabil.*, vol. 25, no. 2, pp. 67–75, Jun. 2010, doi: 10.1177/1088357610364530.

[10] S. P. Orr, L. J. Metzger, and R. K. Pitman, "Psychophysiology of post-traumatic stress disorder," *Psychiatr. Clin. North Am.*, vol. 25, no. 2, pp. 271–293, Jun. 2002, doi: 10.1016/S0193-953X(01)00007-7.

[11] V. Nikouline, J. Ruohonen, and R. J. Ilmoniemi, "The role of the coil click in TMS assessed with simultaneous EEG," *Clin. Neurophysiol.*, vol. 110, no. 8, pp. 1325–1328, Aug. 1999, doi: 10.1016/S1388-2457(99)00070-X.

[12] S. Bestmann, J. Baudewig, H. R. Siebner, J. C. Rothwell, and J. Frahm, "BOLD MRI responses to repetitive TMS over human dorsal premotor cortex," *NeuroImage*, vol. 28, no. 1, pp. 22–29, Oct. 2005, doi: 10.1016/j.neuroimage.2005.05.027.

[13] W. C. Clapp, I. J. Kirk, J. P. Hamm, D. Shepherd, and T. J. Teyler, "Induction of LTP in the human auditory cortex by sensory stimulation," *Eur. J. Neurosci.*, vol. 22, no. 5, pp. 1135–1140, Sep. 2005, doi: 10.1111/j.1460-9568.2005.04293.x.

[14] T. Zaehle, W. C. Clapp, J. P. Hamm, M. Meyer, and I. J. Kirk, "Induction of LTP-like changes in human auditory cortex by rapid auditory stimulation: an fMRI study," *Restor. Neurol. Neurosci.*, vol. 25, no. 3/4, pp. 251–259, May 2007.

[15] E. H. Berger, "Methods of measuring the attenuation of hearing protection devices," *J. Acoust. Soc. Am.*, vol. 79, no. 6, pp. 1655–1687, Jun. 1986, doi: 10.1121/1.393228.

[16] S. P. Schraven, S. K. Plontke, T. Rahne, B. Wasserka, and C. Plewnia, "Hearing safety of long-term treatment with theta burst stimulation," *Brain Stimulat.*, vol. 6, no. 4, pp. 563–568, Jul. 2013, doi: 10.1016/j.brs.2012.10.005.

[17] S. N. Kukke *et al.*, "Hearing safety from single- and double-pulse transcranial magnetic stimulation in children and young adults," *J. Clin. Neurophysiol.*, vol. 34, no. 4, pp. 340–347, Jul. 2017, doi: 10.1097/WNP.0000000000000372.

[18] M. Toivonen, R. Pääkkönen, S. Savolainen, and K. Lehtomäki, "Noise attenuation and proper insertion of earplugs into ear canals," *Ann. Occup. Hyg.*, vol. 46, no. 6, pp. 527–530, Jul. 2002, doi: 10.1093/annhyg/mef065.

[19] R. Neitzel, S. Somers, and N. Seixas, "Variability of real-world hearing protector attenuation measurements," *Ann. Occup. Hyg.*, vol. 50, no. 7, pp. 679–691, Oct. 2006, doi: 10.1093/annhyg/mel025.

[20] A. M. Smith, "Real-world attenuation of foam earplugs," *Aviat. Space Environ. Med.*, vol. 81, no. 7, pp. 696–697, Jul. 2010, doi: 10.3357/ASEM.2817.2010.

[21] H. Nélisse, M.-A. Gaudreau, J. Boutin, J. Voix, and F. Laville, "Measurement of hearing protection devices performance in the workplace during full-shift working operations," *Ann. Occup. Hyg.*, vol. 56, no. 2, pp. 221–232, Mar. 2012, doi: 10.1093/annhyg/mer087.

[22] M. Massimini, F. Ferrarelli, R. Huber, S. K. Esser, H. Singh, and G. Tononi, "Breakdown of cortical effective connectivity during sleep," *Science*, vol. 309, no. 5744, Art. no. 5744, Sep. 2005, doi: 10.1126/science.1117256.

[23] L. M. Koponen, J. O. Nieminen, T. P. Mutanen, M. Stenroos, and R. J. Ilmoniemi, "Coil optimisation for transcranial magnetic stimulation in realistic head geometry," *Brain Stimulat.*, vol. 10, no. 4, pp. 795–805, Jul. 2017, doi: 10.1016/j.brs.2017.04.001.

[24] J. O. Nieminen, L. M. Koponen, and R. J. Ilmoniemi, "Experimental characterization of the electric field distribution induced by TMS devices," *Brain Stimulat.*, vol. 8, no. 3, pp. 582–589, May 2015, doi: 10.1016/j.brs.2015.01.004.

[25] H.-S. Vu and K.-H. Chen, "A low-power broad-bandwidth noise cancellation VLSI circuit design for in-ear headphones," *IEEE Trans. Very Large Scale Integr. VLSI Syst.*, vol. 24, no. 6, pp. 2013–2025, Jun. 2016, doi: 10.1109/TVLSI.2015.2480425.

[26] H.-S. Vu and K.-H. Chen, "Corrections to 'A low-power broad-bandwidth noise cancellation VLSI circuit design for in-ear headphones' [2015 DOI: 10.1109/TVLSI.2015.2480425]," *IEEE Trans. Very Large Scale Integr. VLSI Syst.*, vol. 24, no. 6, pp. 2412–2412, Jun. 2016, doi: 10.1109/TVLSI.2016.2544342.

[27] C. Liu, H. Ding, X. Fang, Z. He, and Z. Wang, "Noise analysis and active noise control strategy of transcranial magnetic stimulation device," *AIP Adv.*, vol. 9, no. 8, p. 085010, Aug. 2019, doi: 10.1063/1.5115522.

[28] T. Paus, P. K. Sipila, and A. P. Strafella, "Synchronization of neuronal activity in the human primary motor cortex by transcranial magnetic stimulation: an EEG study," *J. Neurophysiol.*, vol. 86, no. 4, pp. 1983–1990, Oct. 2001, doi: 10.1152/jn.2001.86.4.1983.

[29] E. M. ter Braack, C. C. de Vos, and M. J. A. M. van Putten, "Masking the auditory evoked potential in TMS–EEG: a comparison of various methods," *Brain Topogr.*, vol. 28, no. 3, pp. 520–528, May 2015, doi: 10.1007/s10548-013-0312-z.

[30] S. J. Schlittmeier, A. Feil, A. Liebl, and J. Hellbrück, "The impact of road traffic noise on cognitive performance in attention-based tasks depends on noise level even within moderate-level ranges," *Noise Health*, vol. 17, no. 76, pp. 148–157, Apr. 2015, doi: 10.4103/1463-1741.155845.

[31] *American Conference of Governmental Industrial Hygienists, Threshold Limit Values and Biological Exposure Indices*. Cincinnati, OH: ACGIH, 2012.

[32] *MIL-STD-1474E. Department of Defense design criteria standard noise limits*. Washington, DC: AMSC 9542, 2015.

[33] *Recommended Practices for Safety and Health Programs*. Washington, DC, USA: Occupational Safety and Health Administration, 2016.

[34] S. M. Goetz, D. L. K. Murphy, and A. V. Peterchev, "Transcranial magnetic stimulation device with reduced acoustic noise," *IEEE Magn. Lett.*, vol. 5, pp. 1–4, Aug. 2014, doi: 10.1109/LMAG.2014.2351776.

[35] A. V. Peterchev, D. L. K. Murphy, and S. M. Goetz, "Quiet transcranial magnetic stimulation: status and future directions," in *2015 37th Annual International Conference of the IEEE Engineering in Medicine and Biology Society (EMBC)*, Aug. 2015, pp. 226–229, doi: 10.1109/EMBC.2015.7318341.

[36] L. M. Koponen, J. O. Nieminen, and R. J. Ilmoniemi, "Minimum-energy coils for transcranial magnetic stimulation: application to focal stimulation," *Brain Stimulat.*, vol. 8, no. 1, pp. 124–134, Jan. 2015, doi: 10.1016/j.brs.2014.10.002.

[37] S. McGee and R. L. McGullough, "Combining rules for predicting the thermoelastic properties of particulate filled polymers, polymers, polyblends, and foams," *Polym. Compos.*, vol. 2, no. 4, pp. 149–161, Oct. 1981, doi: 10.1002/pc.750020403.

[38] Z.-D. Deng, S. H. Lisanby, and A. V. Peterchev, "Electric field depth–focality tradeoff in transcranial magnetic stimulation: Simulation comparison of 50 coil designs," *Brain Stimulat.*, vol. 6, no. 1, pp. 1–13, Jan. 2013, doi: 10.1016/j.brs.2012.02.005.

[39] R. J. Ilmoniemi, "The triangle phantom in magnetoencephalography," *J. Jpn. Biomagn. Bioelectromagn. Soc.*, vol. 22, no. 1, pp. 44–45, May 2009.

[40] M. Kamon, M. J. Tsuk, and J. K. White, "FASTHENRY: a multipole-accelerated 3-D inductance extraction program," *IEEE Trans. Microw. Theory Tech.*, vol. 42, no. 9, pp. 1750–1758, Sep. 1994, doi: 10.1109/22.310584.

[41] J. Starck, I. Rimpiläinen, I. Pyykkö, and T. Esko, "The noise level in magnetic stimulation," *Scand. Audiol.*, vol. 25, no. 4, pp. 223–226, Jan. 1996, doi: 10.3109/01050399609074958.

[42] A. T. Barker, C. W. Garnham, and I. L. Freeston, "Magnetic nerve stimulation: the effect of waveform on efficiency, determination of neural membrane time constants and the measurement of stimulator output.," *Electroencephalogr. Clin. Neurophysiol. Suppl.*, vol. 43, pp. 227–237, 1991.

[43] A. V. Peterchev, S. M. Goetz, G. G. Westin, B. Luber, and S. H. Lisanby, "Pulse width dependence of motor threshold and input–output curve characterized with controllable pulse parameter transcranial magnetic stimulation," *Clin. Neurophysiol.*, vol. 124, no. 7, pp. 1364–1372, Jul. 2013, doi: 10.1016/j.clinph.2013.01.011.

[44] V. Desbeaumes Jodoin, J.-P. Miron, and P. Lespérance, "Safety and efficacy of accelerated repetitive transcranial magnetic stimulation protocol in elderly depressed unipolar and bipolar patients," *Am. J. Geriatr. Psychiatry*, vol. 27, no. 5, pp. 548–558, May 2019, doi: 10.1016/j.jagp.2018.10.019.

[45] J.-P. Miron *et al.*, "Safety, tolerability and effectiveness of a novel 20 Hz rTMS protocol targeting dorsomedial prefrontal cortex in major depression: an open-label case series," *Brain Stimulat.*, vol. 12, no. 5, pp. 1319–1321, Sep. 2019, doi: 10.1016/j.brs.2019.06.020.

[46] R. Ilmoniemi, J. Ruohonen, J. Kamppuri, and J. Virtanen, "Stimulator head and method for attenuating the sound emitted by a stimulator coil," US6503187B1, Jan. 07, 2003.